\documentclass[nofootinbib,aps,twocolumn,pre,a4paper,superscriptaddress,floatfix]{revtex4-1}
\usepackage{amsmath}
\usepackage{graphicx}
\usepackage{natbib}
\usepackage{float}
\usepackage{listings}
\usepackage{color}

\begin{document}
\newcommand{\be}{\begin{equation}}
\newcommand{\ee}{\end{equation}}

\title{General purpose graphics-processing-unit implementation of cosmological domain wall network evolution}
\author{J. R. C. C. C. Correia}
\email{Jose.Correia@astro.up.pt}
\affiliation{Instituto de Astrof\'{\i}sica e Ci\^encias do Espa\c co, Universidade do Porto, CAUP, Rua das Estrelas, 4150-762 Porto, Portugal}
\affiliation{Departamento de F\'{\i}sica, Universidade de Aveiro, Campus Universit\'ario de Santiago, 3810-193 Aveiro, Portugal}
\author{C. J. A. P. Martins}
\email{Carlos.Martins@astro.up.pt}
\affiliation{Instituto de Astrof\'{\i}sica e Ci\^encias do Espa\c co, Universidade do Porto, CAUP, Rua das Estrelas, 4150-762 Porto, Portugal}
\affiliation{Centro de Astrof\'{\i}sica da Universidade do Porto, Rua das Estrelas, 4150-762 Porto, Portugal}
\date{7 September 2017}

\begin{abstract}
Topological defects unavoidably form at symmetry breaking phase transitions in the early Universe. To probe the parameter space of theoretical models and set tighter experimental constraints (exploiting the recent advances in astrophysical observations), one requires more and more demanding simulations, and therefore more hardware resources and computation time. Improving the speed and efficiency of existing codes is essential. Here we present a General Purpose Graphics Processing Unit implementation of the canonical Press-Ryden-Spergel algorithm for the evolution of cosmological domain wall networks. This is ported to the Open Computing Language standard, and as a consequence significant speed-ups are achieved both in 2D and 3D simulations. 
\end{abstract}
\maketitle

\section{Introduction}

Topological defects form at cosmological phase transitions, as a consequence of the Kibble mechanism \cite{Kibble}. If they are stable or sufficiently long-lived they will be present in the more recent universe, as fossil relics of its earlier stages, leading to a plethora of astrophysical signatures \cite{VSH}. The conceptually simplest way of studying the highly nonlinear evolution of defect networks is by thermodynamic analytic modeling. This is based on an idea of Kibble \cite{KIB}, and the current state of the art is the velocity-dependent one-scale model \cite{MS,VOSbook}. However, just as in standard thermodynamics there are parameters (such as Boltzmann's constant) which can not be determined \textit{ab initio} but must be experimentally determined, so analytic models for defect evolution include model parameters which must be determined in high-resolution numerical simulations---thereby calibrating the model.

For the simplest defect model, domain walls from a single scalar field, the canonical approach to field theory simulations is the Press-Ryden-Spergel (PRS) algorithm \cite{PRS}, and state of the art simulations and analytic model calibration are described in \cite{Rybak1,Rybak2}. This WALLS code has been used as a benchmark for Central Processing Units (CPU) and Intel Xeon Phi coprocessors \cite{Intel}. For the more commonly studied case of cosmic strings there are several Goto-Nambu \cite{BB,AS,FRAC,VVO,Blanco} and field theory codes \cite{ABELIAN,Hiramatsu,HindStrings}. There are also implementations for monopoles \cite{Monopoles}, semilocal strings \cite{Semilocals} and non-abelian defects \cite{McGraw,HindNab}. All of these are optimized for standard CPUs, either with shared or distributed memory architectures.

Recent progress in cosmic microwave background \cite{Planck} and gravitational wave detection \cite{GWs} highlights how some of these scenarios can be constrained by high-resolution data. However, they also show that the current bottleneck is the lack of efficient and accurate high-resolution simulations of defect networks that can be used as templates for robust statistical analysis. This will be an even bigger problem for next-generation facilities such as CORE \cite{CORE} and LISA \cite{LISA}: the number and resolution of the required simulations eventually require prohibitive amounts of time or hardware costs. It is therefore important to exploit recent hardware and software advances that yield gains in efficiency of these codes. This work is a step in this direction: we present a first implementation of the PRS algorithm for domain walls on General Purpose Graphical Processing Units (GPGPUs). 

\section{Domain walls and the PRS algorithm}

Domain walls arise whenever a discrete symmetry is broken during a phase transition. The simplest toy model describing wall networks stems from the Lagrangian density of a scalar field $\phi$,
\begin{equation}
    \mathcal{L} = \frac{1}{2}\phi_{,\mu} \phi^{,\mu} - V_0 \left( \frac{\phi^2}{\phi_0^2} - 1 \right)^2\,,
\end{equation}
where the quartic potential $V(\phi)$ has two degenerate minima (and hence the model's vacuum manifold is comprised of two disconnected regions). The equations of motion in a Friedmann-Robertson-Walker universe is obtained by standard variational techniques, leading to
\begin{equation}
    \frac{\partial^2 \phi}{\partial \eta^2} + \alpha \bigg( \frac{d \, ln \, a}{d \, ln \, \eta} \bigg) \frac{1}{\eta} \frac{\partial \phi}{\partial \eta} - \nabla^2 \phi = -\alpha^{\beta} \frac{\partial V}{\partial \phi}\,,
\end{equation}
where $a$ is the scale factor and $\eta$ is the conformal time (related to physical time $t$ by $d\eta = dt / a$). The exact equations of motion have $\alpha=\beta=2$, but one can show that the numerically more convenient case where the walls maintain constant comoving thickness (corresponding to $\beta=0$) still satisfies the appropriate momentum conservation laws provided one simultaneously uses $\alpha=3$. This is the key insight behind the PRS algorithm \cite{PRS}.

This equation can be discretized \cite{PRS} and the evolution of walls is then described by a first-order (with respect to time) Crank-Nicholson, second-order staggered leap-frog scheme, comprised of three different (embarrassingly parallel) steps which in the 2D case can be written
\begin{equation}
    (\nabla^2 \phi)_{i,j} = \phi_{i+1,j} + \phi_{i-1,j}   +\phi_{i, j+1} + \phi_{i,j-1} -4\phi_{i,j}
\end{equation}
\begin{equation}
    \dot{\phi}_{i,j}^{n+1/2} = \frac{(1-\delta)\dot{\phi}_{i,j}^{n-1/2} +\Delta\eta (\nabla^2 \phi_{i,j}^{n} - \partial V / \partial \phi_{i,j}^{n})}{1+\delta}
\end{equation}
\begin{equation}
    \phi_{i,j}^{n+1} = \phi_{i,j}^{n} + \Delta \eta \dot{\phi}_{i,j}^{n+1/2}\,,
\end{equation}
(with a straightforward extension to the 3D case) where the damping term $\delta$ is given by the expression
\begin{equation}
    \delta = \frac{1}{2} \alpha \frac{\Delta \eta}{\eta} \frac{d \, ln \, a}{d \, ln \, \eta}\,.
\end{equation}
In order to characterize wall network evolution, two diagnostic quantities are used. The first is the comoving wall area per unit volume (akin to a density),
\begin{equation}
\rho = \frac{A}{V} = \int \boldsymbol{n} \cdot d \boldsymbol{A} = \Delta A \sum_{links} \delta_\pm \frac{\nabla \phi}{|\phi_{,x}| + |\phi_{,y}|+ |\phi_{,z}|}\,;
\end{equation}
to calculate the area one finds neighboring points where the field changes sign (links), corresponding to an energy concentration associated with the $\phi=0$ local maximum of the field potential. If a link crosses a wall $\delta_\pm$ equals unity, otherwise it vanishes. This has been shown to be a robust method to calculate the area \cite{Ryden}. The second quantity is the square of the product of the average (root-mean squared) velocity $v$ of the network and the corresponding Lorentz factor. This can be calculated from the sum of the ratio between the kinetic and potential energy of each wall (respectively denoted $E_k$ and $V$),
\begin{equation}
    (\gamma v)^2 = \frac{1}{2N} \sum_{walls} \frac{E_k}{V(\phi)}\,,
\end{equation}
where $\gamma=1/\sqrt{1-v^2}$ is the Lorentz factor and the sum is over the number $N$ of grid points containing walls, as identified in the previous step.
 
It is known from analytic arguments \cite{Kibble,VSH}, confirmed with high-resolution simulations \cite{Rybak1,Rybak2} that wall networks in a universe whose scale factor grows as a power law (such as the radiation or matter dominated eras) allowed to evolve for a sufficiently long dynamical range will reach an attractor linear scaling solution which numerically corresponds to
\begin{equation}
    \rho \propto \eta^{\mu}\,,\quad  \gamma v \propto \eta^{\nu},
    \label{eq:12}
\end{equation}
where $\mu=-1$ and $\nu=0$. For simulations with a smaller dynamic range this asymptotic regime may not be reached, which can be identified by a dependence of the exponents $\mu$ and $\nu$ on the box size \cite{Mafalda,Bias}. The purpose of this paper is to present a parallel implementation of the PRS algorithm for 2D and 3D domain walls which runs on GPGPUs, using the behaviour of these two quantities to validate the implementation.

\section{Implementation, speed-ups and error analysis}

Our GPGPU implementation of wall network evolution uses the Open Computing Language (OpenCL) 1.2 framework as specified by the Khronos Consortium \cite{Munshi2012} and implemented by Apple, Inc, and was developed and tested on a machine equipped with a Radeon R9 M395 Graphics Processing Unit, possessing 28 compute units clocked at $834~MHz$, and $2048~MB$ total video memory clocked at $1365~MHz$. On the same machine, the sequential reference version of the same code ran on a Intel i5 6600k with $3.3~Ghz$ core clock (can boost to $3.9~GHz$) and $8192~MB$ of system memory (clocked at $1867~MHz$).

\begin{figure*}
\begin{center}
\includegraphics[width=3.5in]{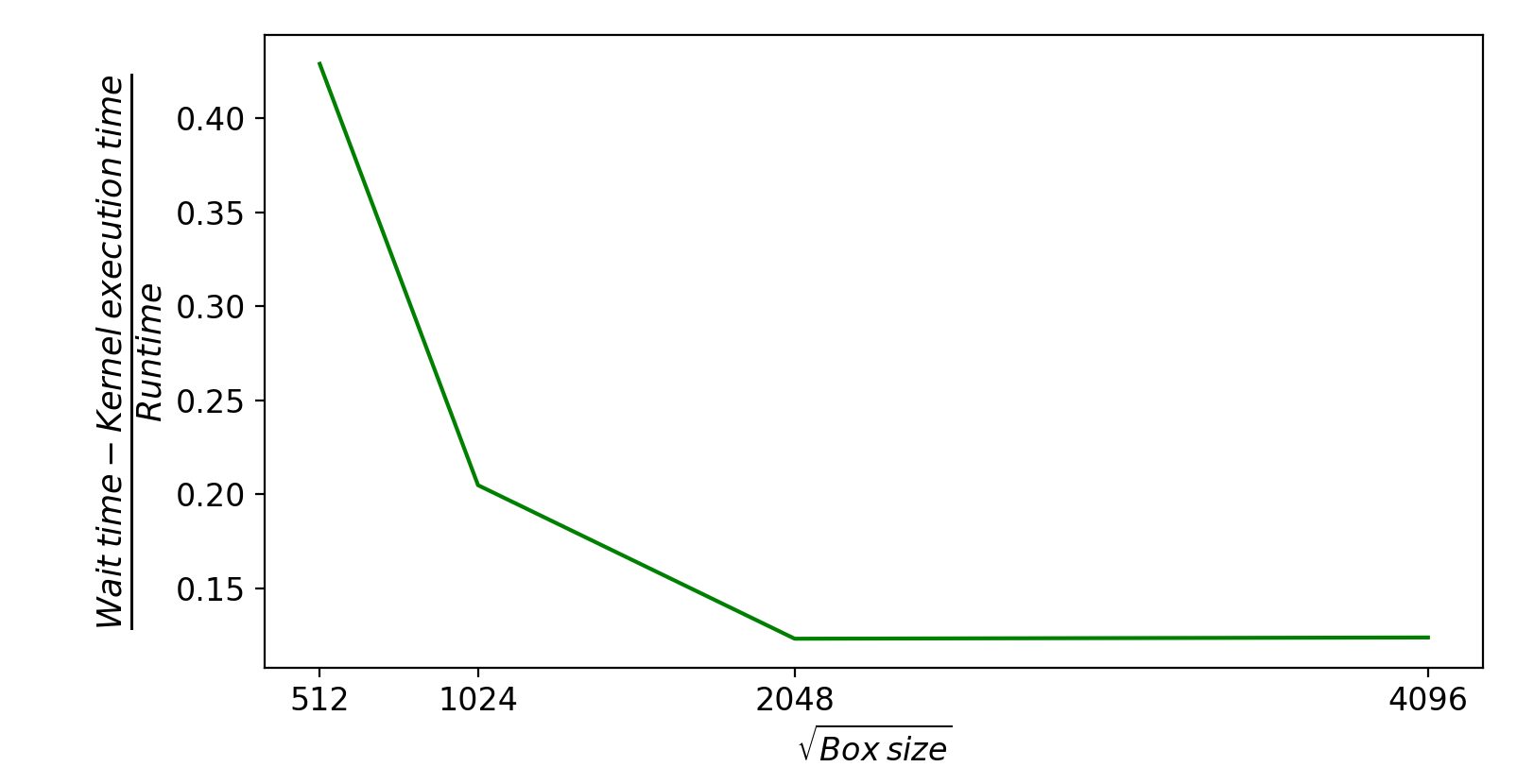}
\includegraphics[width=3.5in]{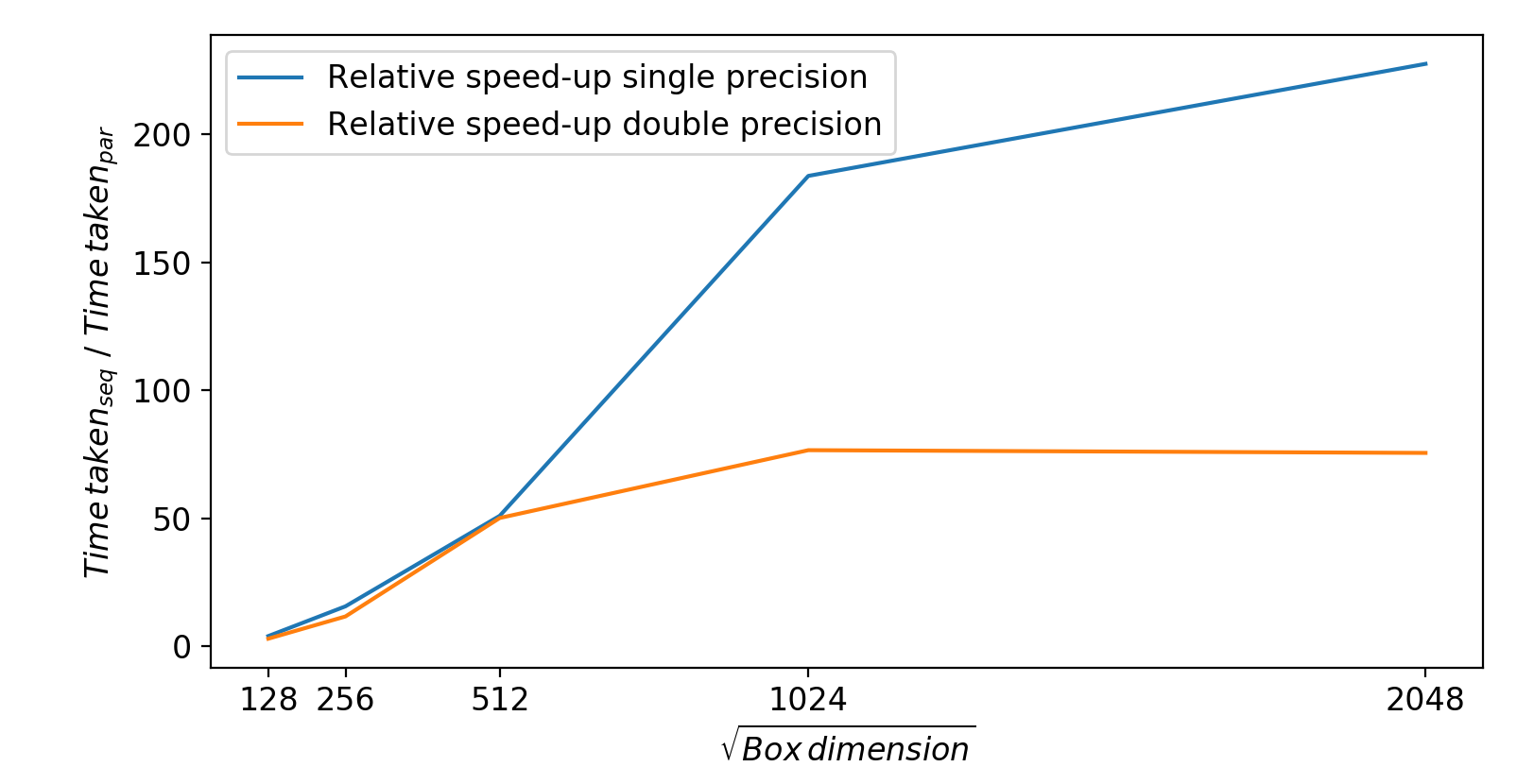}
\includegraphics[width=3.5in]{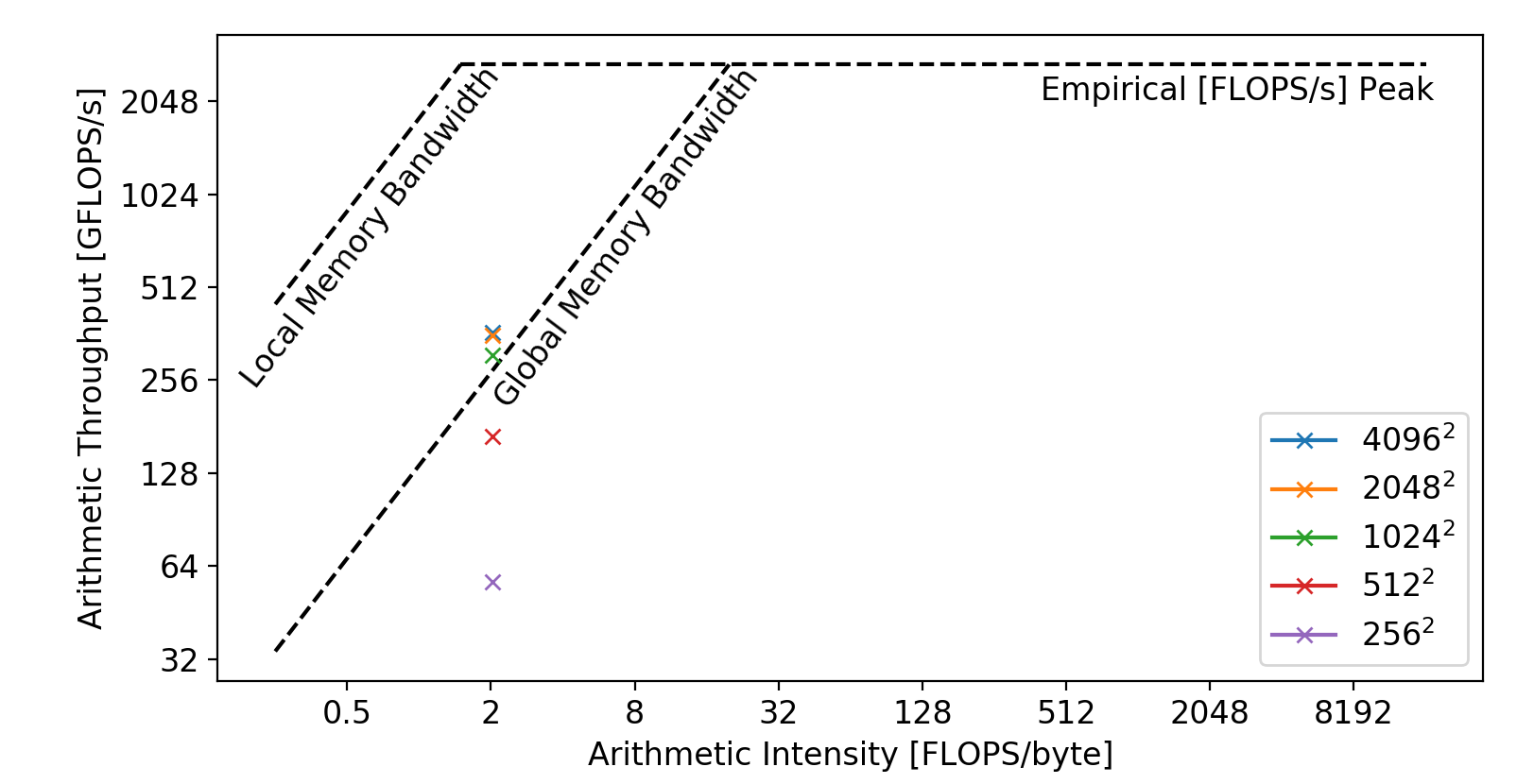}
\includegraphics[width=3.5in]{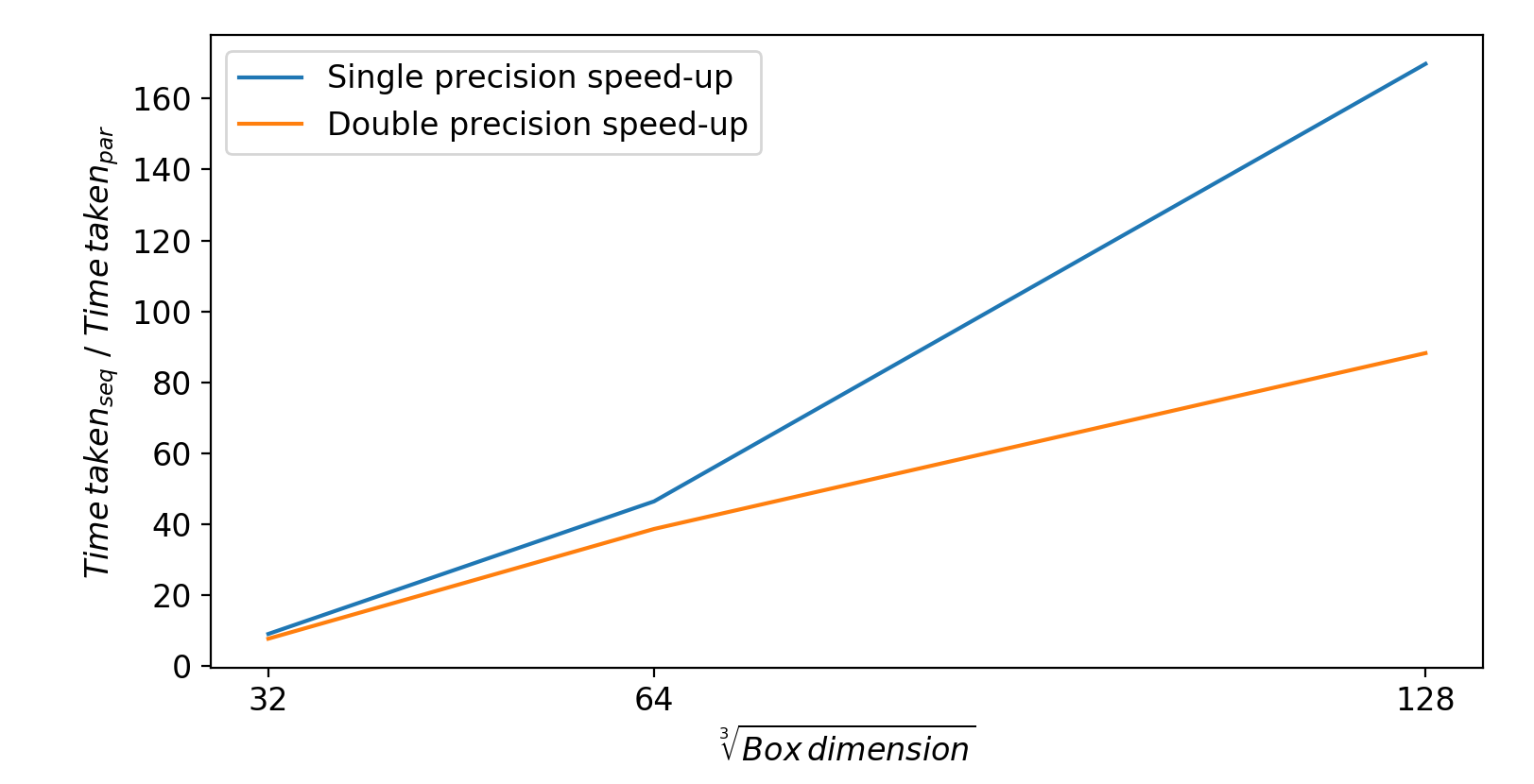}
    \caption{\label{fig1}{\bf Top left:} An estimate of the time wasted in data transfer, or how good the overlap between compute and data transfer is, for different box sizes; {\bf Bottom left:} Roofline model for the 2D implementation; {\bf Right:} relative speed-up of the parallel version when compared to the sequential one, for both single (blue) and double (orange) precision, for 2D (top) and 3D (bottom) simulations.}
\end{center}
\end{figure*}

In OpenCL, applications are subdivided in data-parallel functions named kernels, which are to be compiled at run-time (Just In-Time compilation). Each step of the PRS algorithm corresponds to a kernel, and so do the velocity and density calculations, with a separate kernel for the sums. These kernels execute in order, one timestep at a time. For the sum reduction kernel, we use the scalar version of the kernel in \cite{Scarpino2011}. The reason for not using the vector one (where instead of using vector data-types like \textit{float4}, one would use \textit{float}, for instance), is that the preferred  vector width\footnote{The OpenCL compiler automatically packs the preferred number of work-items or threads into Single-Instruction-Multiple-Data lanes and henceforth takes advantage of the native vector width. The native width is the number of elements a Vector Arithmetic Logic Unit can process at once.} of the device in question is, for both double and floating point types
\begin{lstlisting}
CL_DEVICE_NATIVE_VECTOR_WIDTH_FLOAT: 1
CL_DEVICE_PREFERRED_VECTOR_WIDTH_FLOAT: 1
CL_DEVICE_NATIVE_VECTOR_WIDTH_DOUBLE: 1
CL_DEVICE_PREFERRED_VECTOR_WIDTH_DOUBLE: 1
\end{lstlisting}
so it is equivalent to use either kernel. The sum reduction kernel computes a partial sum for each local memory\footnote{The OpenCL memory model describes several types of memory: Global (which on a graphics card corresponds to video memory), local (a fast-access cache on each compute unit), constant (technically part of video memory as well, but constant) and private (memory bound to each work-item/thread). There is a tendency in this code to try to utilize local memory whenever possible, due to its faster access times. We note that we still need to port two kernels to use local memory: the Laplacian and the density kernel. Concatenation of kernels should also follow suit, in order to reduce the number of times one copies to and from memory.} patch, and all partial sums are transferred back to the host side, summed and written to disk. The only role of the CPU is to sum the partial sums which result from the calculation of the velocity and the density As a small optimization we use two queues running asynchronously with respect to each other, ensuring overlap between execution of compute kernels and data transfer operations.

\begin{figure*}
\begin{center}
\includegraphics[width=3.5in]{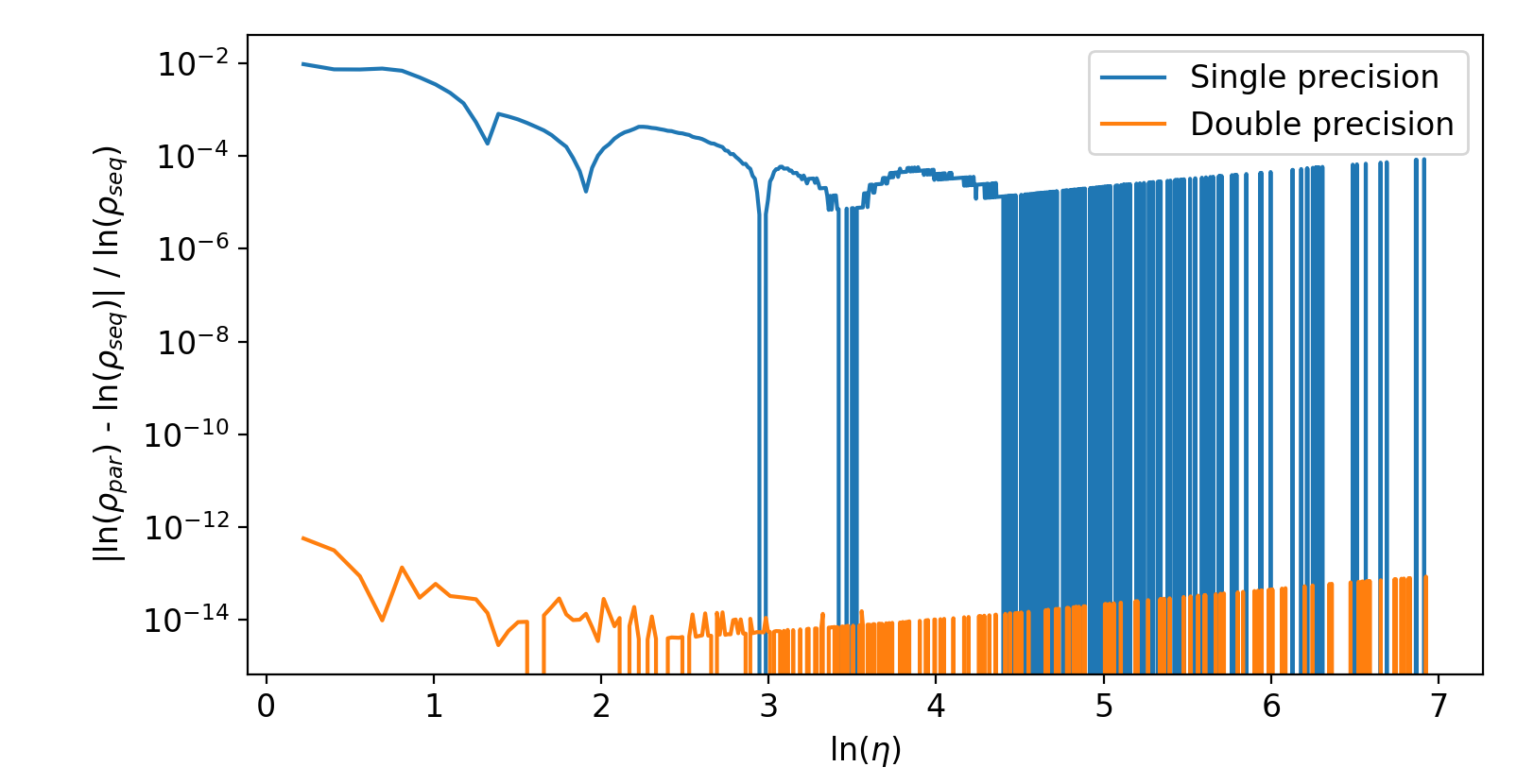}
\includegraphics[width=3.5in]{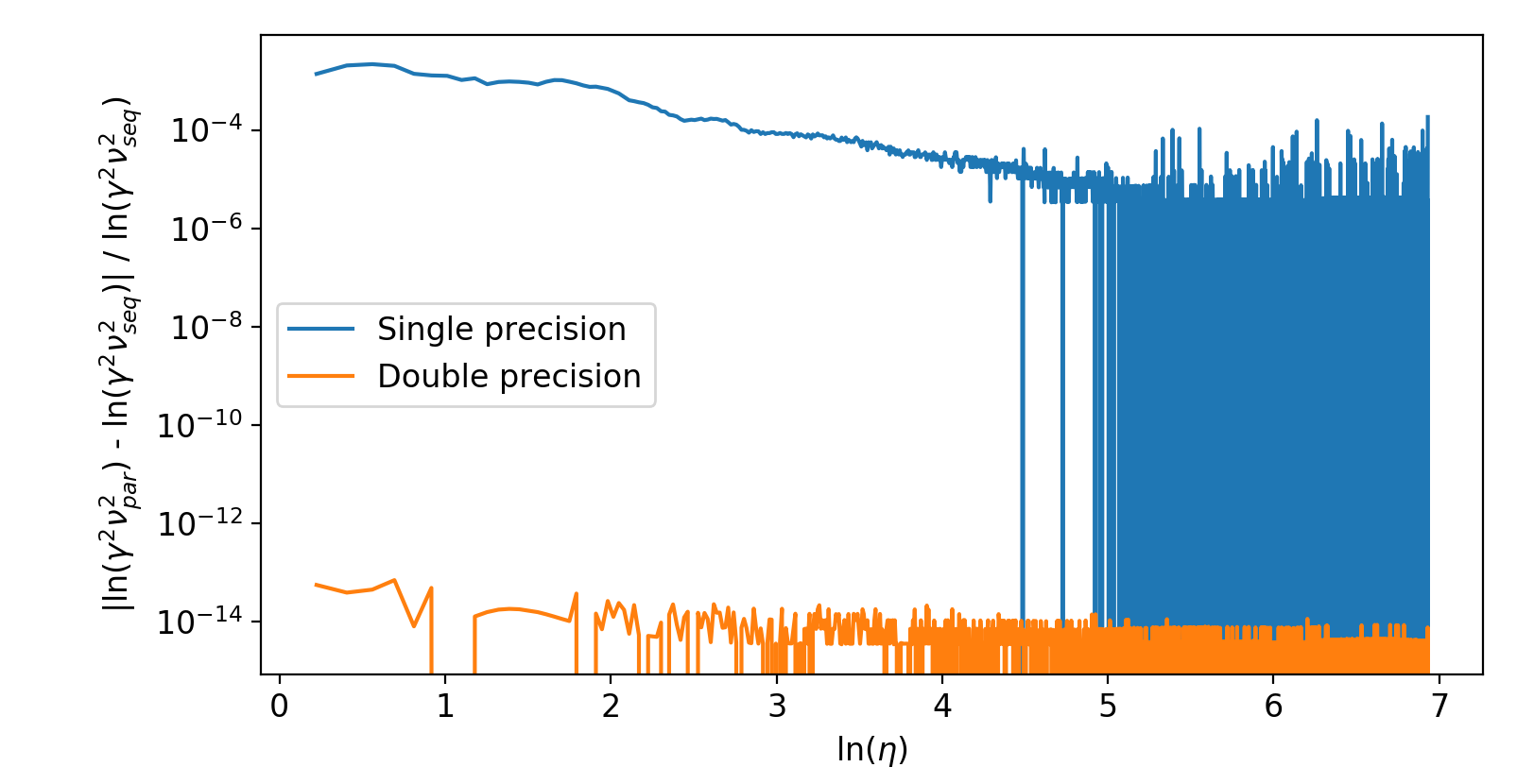}
\includegraphics[width=3.5in]{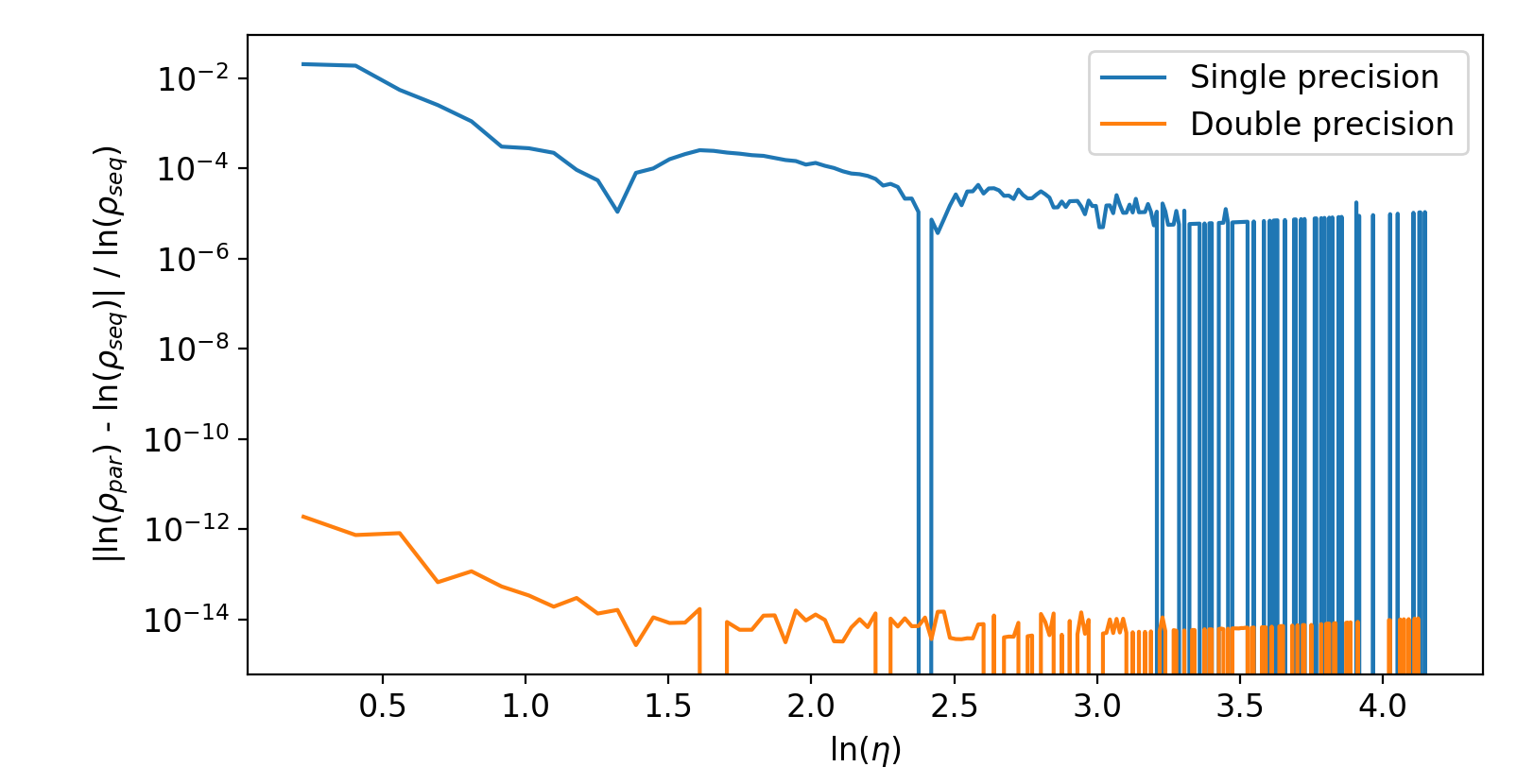}
\includegraphics[width=3.5in]{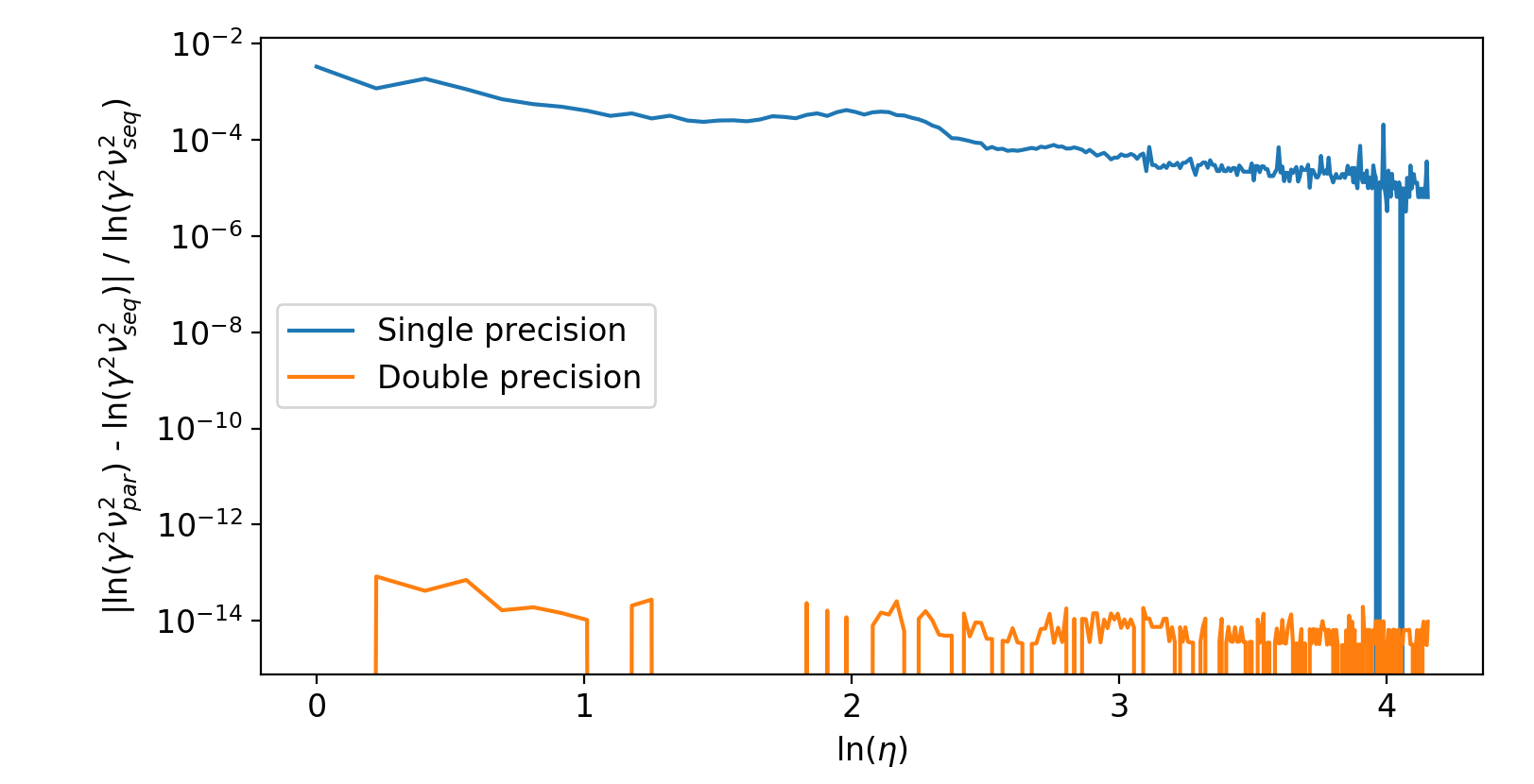}
\caption{\label{fig2}Relative error between sequential and parallel code implementations, with $2048^2$ boxes (top panels) and $128^3$ (bottom panels), for both single (blue) and double precision (orange), for the wall density (left panels) and the velocity (right panels).}
\end{center}
\end{figure*}

Our code is compatible with both double and single precision, though it should be noted that consumer facing graphics cards usually have much lower peak operations per second and as such there is a severe speed penalty in utilizing double precision (for AMD cards based on the Graphics Core Next architecture this varies between $1/2$ and $1/16$ of peak single precision operations per second \cite{AMD2012,AMD2014}). This expectation is confirmed by our analysis, summarized in Fig. \ref{fig1}, which also highlights the large relative speed-up of going parallel provided the box size is large enough to fully exploit a GPU. 

We also note a few characteristics of the implementation. The fields are represented in memory using buffer data (linear contiguous), and the number of threads (work-items) spawned are always equal to the number of points in a box. The OpenCL compiler (and the underlying hardware) handle the distribution of threads automatically. The implementation has low arithmetic intensity, and seems mostly compute bound (when taking local memory bandwidth into account, see roofline model in Fig. \ref{fig1}). From AMD's CodeXL, we report that all kernels have an occupancy of 70\% and the main bottleneck on the number of waves per SIMD unit seems to be the number of scalar registers (96 are used, which corresponds to a score of 8/10, below 81 would be ideal). The tool also shows that the implementation would highly benefit from more local memory and more vector register usage (where 4-23 vector registers are used, depending on the kernel). A prime example of a kernel which could still benefit from local memory usage would be the density kernel (25.81\% of runtime, the most time-consuming kernel), where locality could be a way to tile memory. This is not to say that we don't already employ local memory in some places, examples include the velocity kernel (where we highlight the increased granularity of atomic additions needed to count the number of walls, as seen in \cite{stream}).

In order to quantify if there is a data transfer bottleneck, we first remark how the overlap between compute and data transfer works: one has two different queues, one for data transfer, one for kernel execution, and using events one triggers data transfer upon completion of the partial sums kernel. Unfortunately, to allow for overlap, the enqueueing of data transfers needs to be non-blocking. After enqueueing some kernels, it is important to wait for the data transfers to complete (to ensure the sum of partial sums isn't summing over garbage). Since the waiting time will also include waiting for compute kernels to finish (again enqueueing kernels is a non-blocking operation) we estimate the time taken by data transfer to roughly correspond to the difference between waiting time and total kernel execution time. Comparing to the runtime reveals that data transfer is only a bottleneck in low resolution boxes.

Significant loss of precision need not occur from single precision, though in OpenCL division and square root operations do not generally apply correct rounding as prescribed by the IEEE754 specification. For this graphics card (and implementation), a JIT compiler flag can enable it by passing the option \textit{-cl-fp32-correctly-rounded-divide-sqrt} for single precision arithmetic only. Since this option is not available for double precision, we compare the sequential and parallel implementations for the two aforementioned diagnostic quantities, either using single or double precision. To do so we evolve several boxes with the described settings using the same initial conditions across the board (generated by the single precision code, to avoid hamstringing the single precision version at initial time-steps due to type-casting rounding errors). 

In both the double and the single precision case, the differences between the parallel and sequential versions seem to be negligible after the early timesteps, once the wall networks have eased the 'numerical' initial conditions in the box and are approaching the scaling solution. Both errors seem bound by the maximum precision specified by their data-types (for this specific machine and as dictated by the \textit{FLT\_DIG} and \textit{DBL\_DIG} macros) at latter timesteps. This can be seen in Fig. \ref{fig2}. Note that the single precision case tends to incur a much larger error during the initial timesteps.

\begin{figure*}
\begin{center}
\includegraphics[width=3.5in]{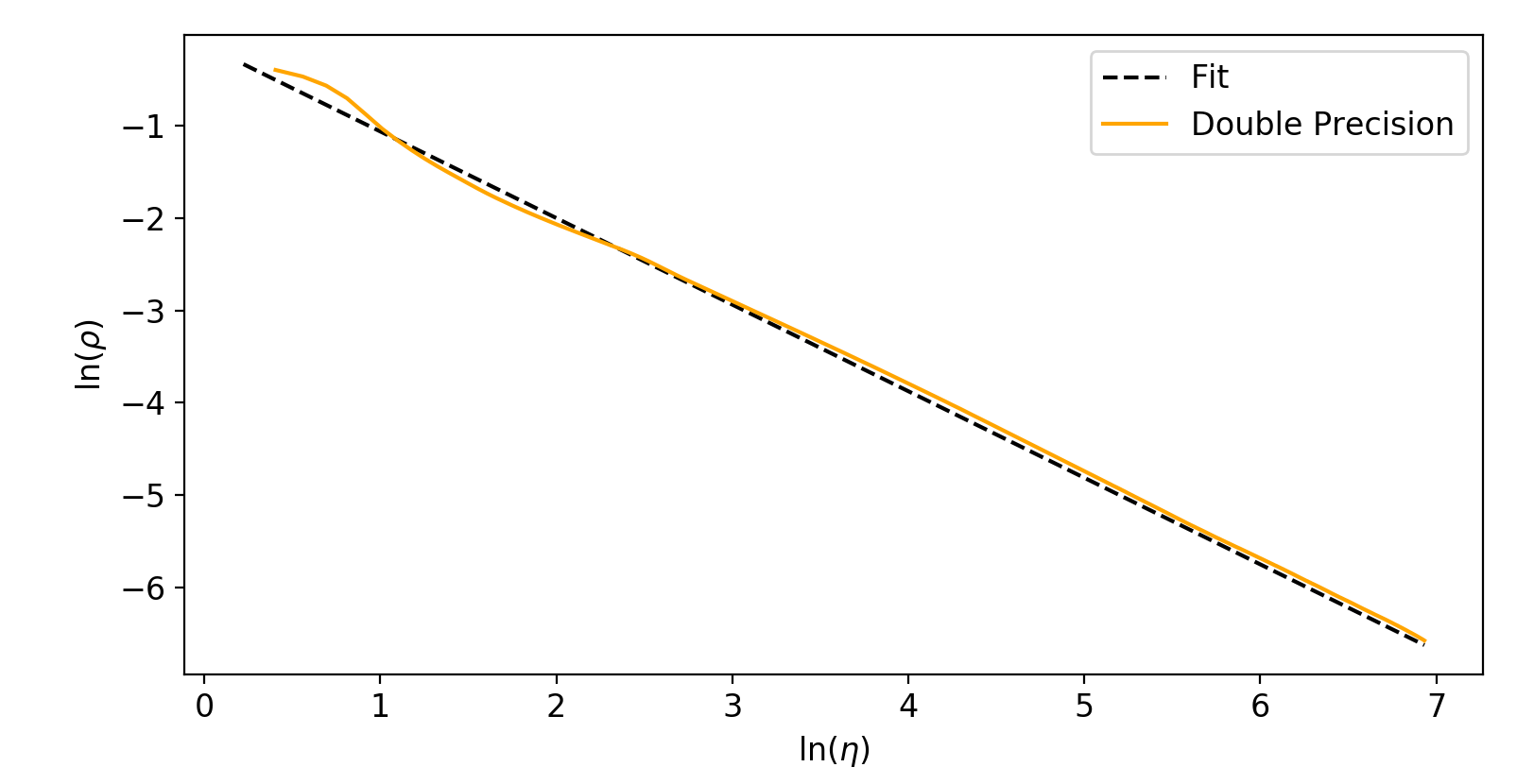}
\includegraphics[width=3.5in]{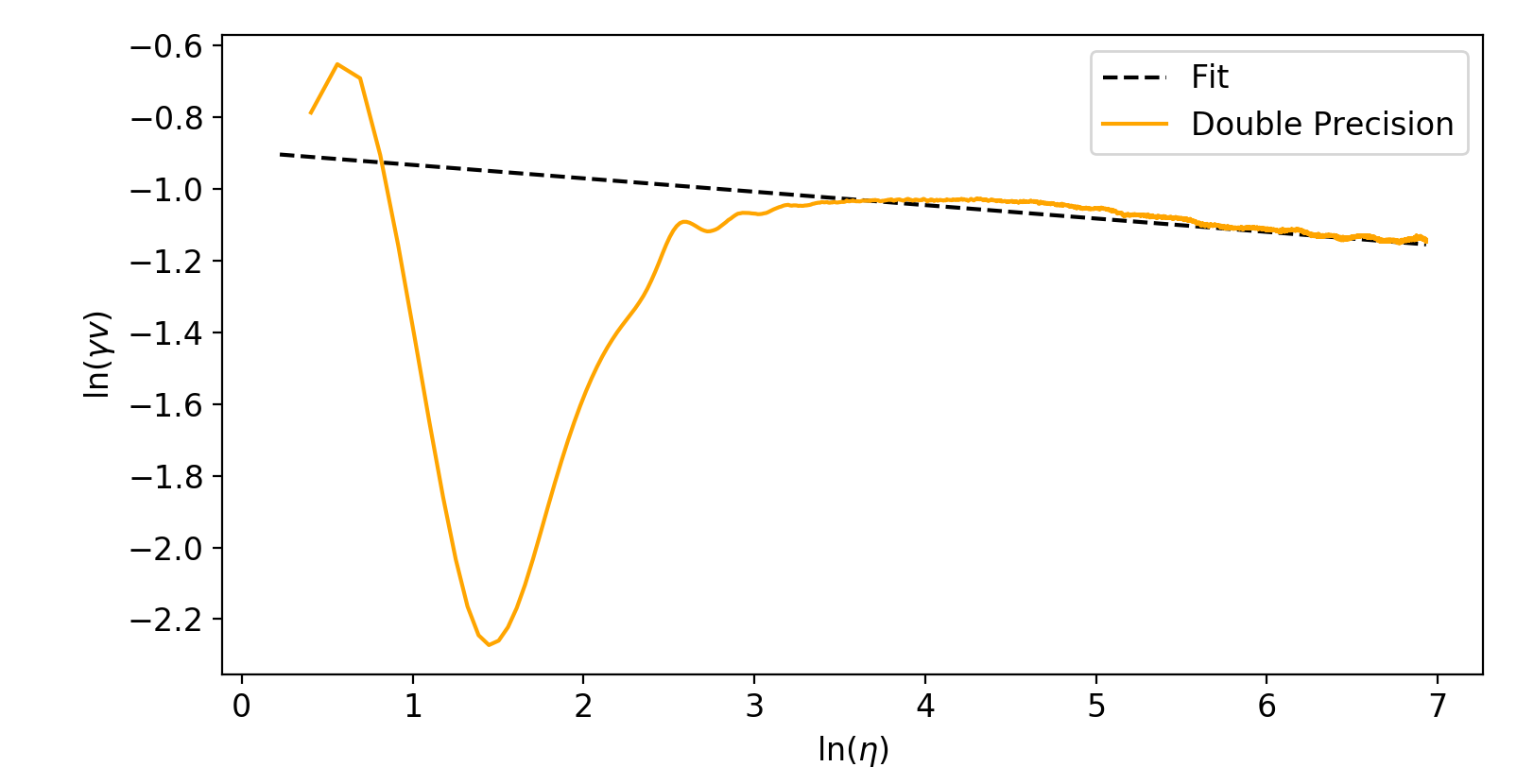}
\includegraphics[width=3.5in]{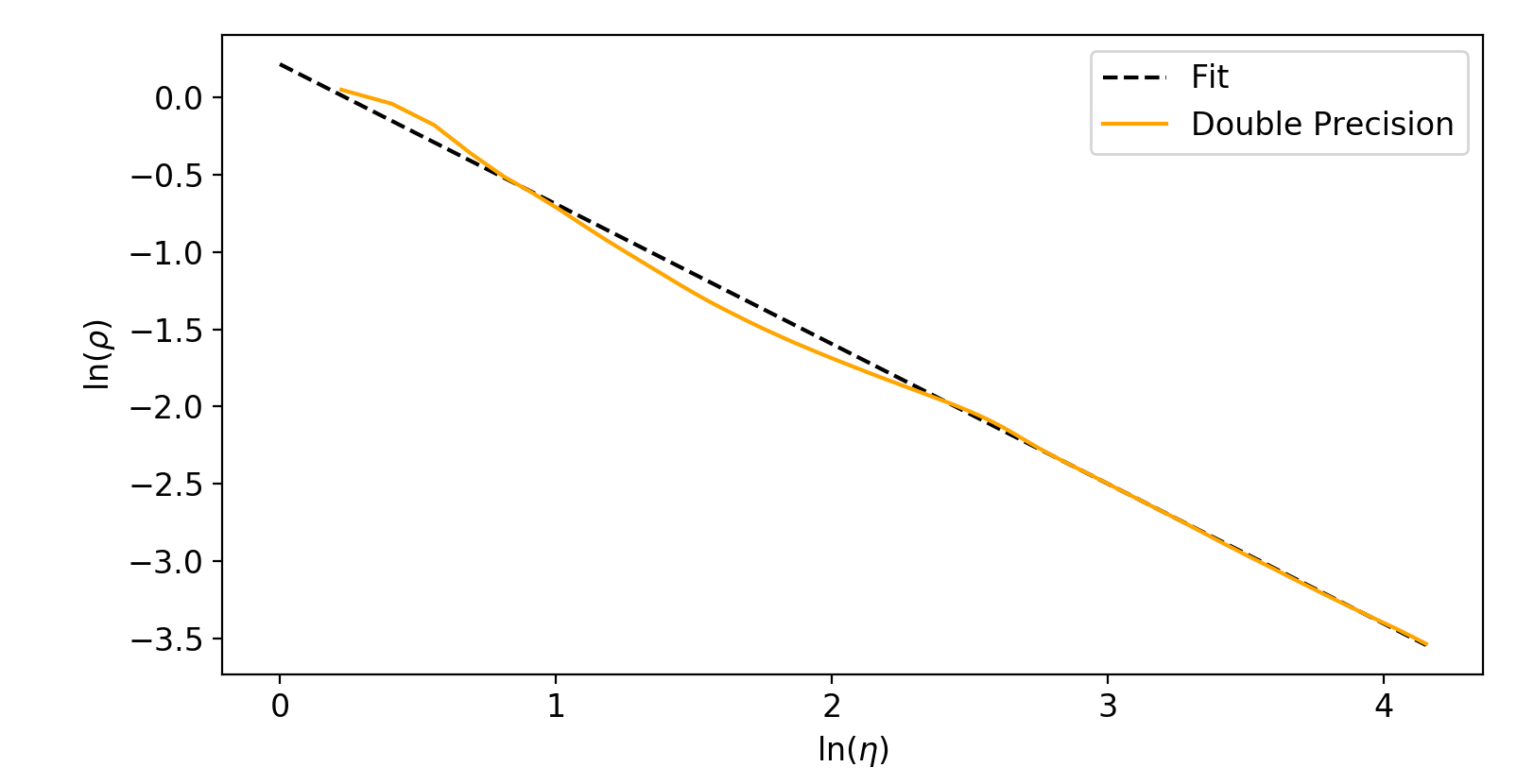}
\includegraphics[width=3.5in]{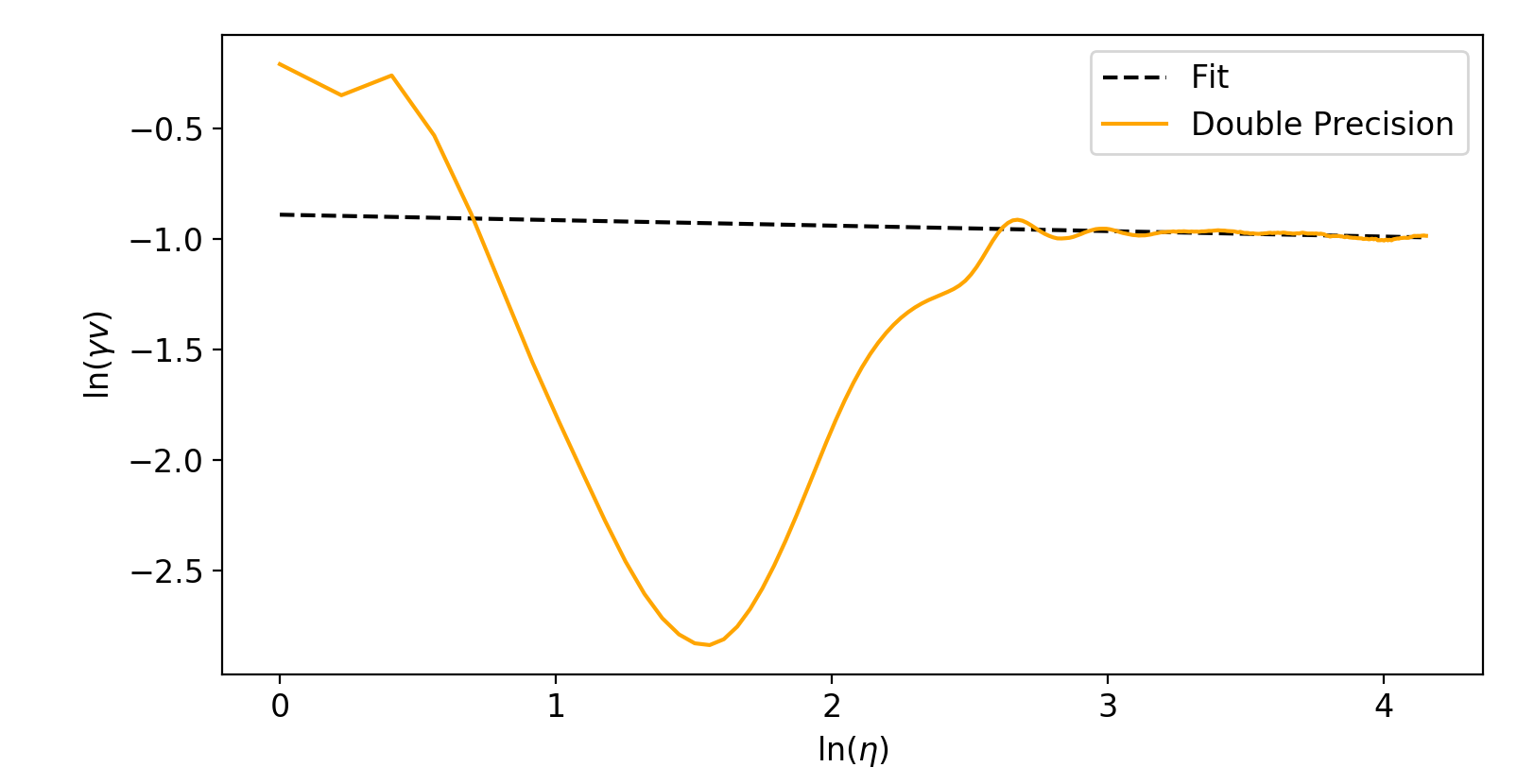}
\caption{\label{fig3}Evolution of the density ($\rho$, left panels) and the velocity ($\gamma v$, right panels), for $2048^2$ and $128^3$ box simulations (top and bottom panels, respectively), showcasing the expected scaling behaviour.}
\end{center}
\end{figure*}

As a final validation, we use sets of five single and double precision runs of $2048^2$ and $128^3$ boxes to calculate the scaling exponents defined in Eq. \ref{eq:12}. The same set of five fixed seeds is used both in single and in double precision. The scaling exponents are calculated using a linear fit and ignoring the early part of the simulations (whose dynamics is still dominated by the initial conditions). The calculated exponents are listed in Table \ref{tab:scal}, and are in agreement with previous simulations of boxes of these sizes with CPU versions of the code \cite{Mafalda,Bias}. The listed uncertainties are statistical, from the average of each set of five runs (this is the relevant comparison here); additional systematic uncertainties in these diagnostics are discussed in \cite{Rybak2}. Figure \ref{fig3} depicts the evolution of the density and the velocity, illustrating the expected approach to the scaling behaviour.

\begin{table}
\caption{\label{tab:scal}Scaling exponents $\mu$ and $\nu$ (with 1$\sigma$ statistical errors) for single and double precision runs, calculated using the points beyond $\log(\eta)=2.58$ for both $2048^2$ and $128^3$ simulations.}
\begin{center}
    \begin{tabular}{ c | c | c }
\hline
$2048^2$ & $\mu$ & $\nu$ \\
\hline
Single precision & $-0.9381 \pm 0.0003$ & $-0.0374 \pm 0.0005$ \\ 
Double precision & $-0.9381 \pm 0.0003$ & $-0.0374 \pm 0.0005$ \\ 
\hline
\hline
$128^3$ & $\mu$ & $\nu$ \\
\hline
Single precision & $-0.956 \pm 0.003$ & $-0.034 \pm 0.006$ \\ 
Double precision & $-0.905 \pm 0.002$ & $-0.025 \pm 0.004$ \\ 
\hline
    \end{tabular}
\end{center}
\end{table}

\section{Conclusion}

We have ported a previous sequential code based on the PRS algorithm to a parallel OpenCL-based implementation, specifically optimized to the GPU used. This highlights the point that even with a consumer grade graphics card reasonable speed-ups are to be expected, provided a large enough box size is used. We also investigated the possible loss of precision. The fastest version but with higher error corresponds to the single precision version with compiler flag \textit{-cl-fp32-correctly-rounded-divide-sqrt}. Both the single and double precision version yield consistent results for the scaling diagnostics (keeping in mind that larger boxes yield better results). 

The bottleneck at larger box sizes will be the amount of memory available to the graphics card; this might be lessened by reducing memory usage. Changing graphics card will require re-optimization of the code. Further ongoing work includes optimization for Central Processing Units (as OpenCL guarantees portability of code---minor implementation differences aside---but not optimized execution for all types of devices) and a comparison between the parallel codes on the GPU and on the CPU. After these further validations and verifications we expect that the GPU domain wall codes may be used for generating large sets of production runs for astrophysical exploitation. 

\begin{acknowledgments}
This work was done in the context of project PTDC/FIS/111725/2009 (FCT, Portugal), with additional support from grant UID/FIS/04434/2013. CJM is supported by an FCT Research Professorship, contract reference IF/00064/2012, funded by FCT/MCTES (Portugal) and POPH/FSE (EC).
\end{acknowledgments}

\bibliography{artigo}
\end{document}